\documentclass[12pt]{article}
\usepackage{graphicx}
\usepackage{amsfonts}
\usepackage{amssymb}
\newcommand{\nin}{\noindent}
\newcommand{\be}{\begin{equation}}
\newcommand{\ee}{\end{equation}}
\newcommand{\bea}{\begin{eqnarray}}
\newcommand{\eea}{\end{eqnarray}}
\newcommand{\br}{\hskip .25cm/\hskip -.25cm}

\newcommand{\nn}{\nonumber\\}

\newcommand{\ol}{\overline}

\begin{document}
\begin{flushleft}
KCL-PH-TH/2011-22
\end{flushleft}
\vspace{0.5cm}
\begin{center}

{\bf{\Large Dynamical mass generation in Lorentz-violating $QED$}}

\vspace{1cm}

{\bf J.~Alexandre}\\
jean.alexandre@kcl.ac.uk\\
King's College London, Department of Physics, London WC2R 2LS, UK\\

\vspace{1cm}

{\bf Abstract}

\end{center}

A Lorentz violating modification of massless $QED$ is proposed, with higher order space derivatives for the photon field. 
The fermion dynamical mass generation is studied with the Schwinger-Dyson approach, and it is found that
the resulting mass is many orders of magnitude 
smaller than the (Plank-) mass scale introduced by the higher order derivative terms.
This is due to a suppressing factor,
non-analytic in the fine structure constant and exponentially small. This scenario is an 
alternative to the Higgs mechanism for the electron mass, which arises here from Lorentz violating effects at high energies.

\eject

Recently, field theories with higher-order space derivatives have attracted attention, because of the improvement 
of graph convergence 
\cite{lifshitz}, at the price of violating Lorentz symmetry at high energies. 
Ghosts are not introduced by this procedure, since the time derivative order remains minimal, such that 
no new pole appears in the propagator of particles.
Also, theories based on anisotropic scaling of space and time (Lifshitz type theories) allow new 
renormalizable interactions, and for
example a renormalizable exponential potential, in 3+1 dimensions, has been studied in \cite{LL}. Finally, 
a renormalizable Lifshitz type theory of 
Gravity has been proposed, which could lead to Quantum Gravity \cite{Horava}.\\
A Lorentz-violating extension of massless Quantum Electrodynamics ($QED$) is discussed here, 
where higher order space derivatives are introduced for the photon field. 
This model has isotropic scaling in space time though, such that the higher order space derivatives involve a mass scale $M$, whose
role will be discussed a bit further. Higher order space derivatives were  already considered in a modified 
Dirac equation \cite{Pavlopoulos}, where the resulting phenomenology concerning gamma ray bursts is also discussed.
Higher order space derivatives could arise, for example, from quantum gravitational space time foam \cite{foam}.
The mass $M$ (which has to thought of as the Plank mass) naively suppresses the effect of higher order derivatives in the IR,  
but as we will see, an IR signature of these higher orders remains: a fermion mass is generated dynamically. 
This mass, although proportional to $M$, is orders of magnitude smaller than $M$ 
since it is suppressed by an exponentially small function of the fine structure constant.
We will study here this mass generation, using the non-perturbative Schwinger-Dyson approach, and we will discuss 
the properties of the result, which can be an alternative to the Higgs mechanism. 
We note here that another alternative to the Higgs model has been proposed in the context of anisotropic theories \cite{nohiggs},
and dynamical mass generation has been studied in \cite{fourfermion} for a Lifshitz-type
four-fermion model, and in \cite{yukawa} for a Lifshitz type Yukawa model.\\
Finally, it is important to mention the first works on dynamical mass generation in $QED$
\cite{johnson}, where a relation between the bare and dressed masses is derived, which involves a cut off $\Lambda$. In this
context, it was shown that, in order to have a finite theory, the limit $\Lambda\to\infty$ implies that the bare mass 
should vanish, such that the dressed mass must be of dynamical origin. The Schwinger Dyson equation 
with a finite cut off was studied later \cite{lambda}, where a critical value for the fine structure constant was found
in order to have dynamical mass generation in the limit $\Lambda\to\infty$. A summary of these results can be found in 
\cite{miransky}, and we stress here that, in the present work,
$M$ is not a cut off and is not meant to be sent to infinity.

The Lorentz-violating Lagrangian considered here is 
\be\label{bare}
{\cal L}=-\frac{1}{4}F^{\mu\nu}\left(1-\frac{\Delta}{M^2}\right)F_{\mu\nu}
-\frac{\xi}{2}\partial_\mu A^\mu\left(1-\frac{\Delta}{M^2}\right)\partial_\nu A^\nu 
+i\ol\psi\br D\psi,
\ee
where $D_\mu=\partial_\mu+ieA_\mu$, and $\Delta=-\partial_i\partial^i=\vec\partial\cdot\vec\partial$ 
(the metric used is (1,-1,-1,-1)),
which recovers $QED$ in a covariant gauge if $M\to\infty$. 
The Lorentz-violating terms have two roles: introduce a mass scale, necessary to generate a fermion mass, 
and lead to finite gap equation, as will be seen further. Note that, in the studies of dynamical mass generation 
in $QED$ in the presence of an external magnetic field $B$ (``magnetic catalysis'' \cite{mag}), 
the gap equation, in the Lowest Landau Level approximation, is finite because of the mass scale $\sqrt{|eB|}$ 
- where $e$ is the electric charge - which plays the role of a gauge invariant cut off.
We stress here that $M$ is not the regulator of the theory (\ref{bare}), since it regularizes loops 
with an internal photon line only. $M$ is rather a parameter of the model, which, if of the order of the Plank mass, 
cannot have realistically measurable consequences, besides the fermion dynamical mass derived in this study.
Also, the Lorentz violating modifications proposed in the Lagrangian (\ref{bare}) do not 
alter the photon dispersion relation, which remains relativistic.
The one-loop polarization tensor is the same as the one calculated in usual $QED$, 
since it depends on the fermion propagator only (this graph should include the dynamical mass, thought, 
such that it is actually not a one-loop graph, but contains a partial resumation already). As 
a consequence, the ``one-loop'' (with the dynamical mass) running coupling constant of the model (\ref{bare})
is the same as the one calculated in usual $QED$: because of gauge invariance, the fermion wave function renormalization is the same 
as the vertex correction (for vanishing incoming momentum), and only the polarization tensor enters 
into account for the corrections to the coupling. 

Finally, no higher order space derivatives are introduced for the fermions, for the following reason: 
in order to respect gauge invariance, such terms would need to be of the form
\be
\frac{1}{M^{n-1}}\ol\psi (i\vec D\cdot\vec\gamma)^n\psi~~~~~~n\ge 2,
\ee
such that new and non-renormalizable couplings would be introduced.\\
From the Lagrangian (\ref{bare}), the bare photon propagator is given by
\be\label{D}
D_{\mu\nu}^{bare}(\omega,\vec p)=\frac{i}{1+p^2/M^2}\left( \frac{\eta_{\mu\nu}}{-\omega^2+p^2}-
\zeta\frac{p_\mu p_\nu}{(\omega^2-p^2)^2}\right) ,
\ee
where $\zeta=1/\xi-1$, $p_0=\omega$ and $p^2=\vec p\cdot\vec p$.
The Schwinger-Dyson equation for the fermion propagator is \cite{miransky}:
\be\label{SD}
G^{-1}-G_{bare}^{-1}=\int D_{\mu\nu}(e\gamma^\mu) G\Gamma^\nu,
\ee
where $\Gamma^\nu$, $G$ and $D_{\mu\nu}$ are respectively the dressed vertex, the dressed fermion propagator 
and the dressed photon propagator. This equation gives an exact self consistent relation between {\it dressed} 
$n$-point functions, and thus is not perturbative.
As a consequence, no redefinition of bare parameters can be done in order to absorb would-be divergences, 
and for this reason one needs this equation to be regularized by $M$, which will have a physical meaning. \\

In order to solve the Schwinger-Dyson equation (\ref{SD}), we consider 
the so-called ladder (or rainbow) approximation, which consists in neglecting corrections to the vertex. 
It is known that this approximation is not gauge invariant \cite{miransky}, but, as we will see, 
the gauge coupling dependence of the dynamical mass is not affected by the choice of the gauge parameter $\xi$.
Then we will neglect corrections to the photon propagator, as well as the fermion wave function renormalization: 
only the fermion dynamical mass will be taken into account as a correction, such that the dressed fermion propagator 
will be taken as
\be\label{G}
G(\omega,\vec p)=i\frac{\omega\gamma^0-\vec p\cdot\vec\gamma+m_{dyn}}{\omega^2-p^2-m_{dyn}^2},
\ee
where $m_{dyn}$ is the fermion dynamical mass.
With these approximations, the Schwinger-Dyson equation (\ref{SD}) - involving a convergent integral - leads to 
\be
m_{dyn}=\frac{\alpha}{\pi^2}\int_{-\infty}^\infty d\omega\int_0^\infty
\frac{p^2dp}{1+p^2/M^2}\frac{m_{dyn}(4+\zeta)}{(\omega^2+p^2)(\omega^2+p^2+m^2_{dyn})},
\ee
where the fine structure constant is $\alpha=e^2/4\pi$.
This equation has the obvious solution $m_{dyn}=0$, and potentially a second solution, which must
satisfy the following gap equation, obtained after integration over the frequency $\omega$,
\be
\frac{\pi}{(4+\zeta)\alpha}=\int_0^\infty\frac{xdx}{1+\mu^2x^2}\left( 1-\frac{x}{\sqrt{1+x^2}}\right),
\ee
where $\mu=m_{dyn}/M$ is the dimensionless dynamical mass, expected to be very small $\mu<<1$, and $x=p/m_{dyn}$.
Both terms in the last equation, if taken separately, lead to diverging integrals, with divergences which cancel each other.
An integration by parts for the second term leads then to
\bea\label{8}
\frac{2\pi}{(4+\zeta)\alpha}&=&\frac{1}{\mu^2}\int_0^\infty dx\frac{\ln(1+\mu^2x^2)}{(1+x^2)^{3/2}}\nn
&\simeq&\frac{1}{\mu^2}\int_1^{1/\mu}\frac{\mu^2x^2}{x^3}=\ln\left( \frac{1}{\mu}\right)
\eea
and the fermion dynamical mass is finally given by
\be\label{mdyn}
m_{dyn}\simeq M\exp\left(-\frac{2\pi}{(4+\zeta)\alpha}\right).
\ee
Note that a numerical analysis could be performed in order to solve eq.(\ref{8}), before the expansion for small $\mu$, 
but might not emphasize the non-analytic
feature of the dynamical mass as well as the expression (\ref{mdyn}).\\

To conclude, we discuss the main result of this study.\\
{\it (i)} Among the two solutions $m_{dyn}=0$ and $m_{dyn}\ne0$, the physical system chooses the non-vanishing dynamical mass,
in order to avoid IR instabilities, not favorable energetically, which would otherwise occur in the theory; \\
{\it (ii)} The expression (\ref{mdyn}) for $m_{dyn}$ is not analytic in $\alpha$, 
such that a perturbative expansion cannot lead to such a result, which
justifies the use of a non-perturbative approach. A perturbative expansion would lead to the solution $m_{dyn}=0$ only;\\
{\it (iii)} There is an obvious dependence on the gauge parameter $\zeta$, which has a consequence on the value of $m_{dyn}$. But 
the important point is the non-analytic $\alpha$-dependence of the dynamical mass, which is not affected by the choice of gauge. 
To eliminate this gauge dependence, one should consider better approximations that the ones used here to solve the Schwinger Dyson 
equation (\ref{SD}), but the resulting dynamical mass would still
be of the form $M\exp(-c/\alpha)$, where $c$ is a constant of order 1. 
This feature is known in the studies of dynamical mass generation 
in $QED$ in the presence of an external magnetic field \cite{mag};\\
{\it(iv)} Even for a very large mass scale $M$, the suppressing exponential factor leads to a small dynamical mass. 
Indeed, if we consider the electron mass $m_{dyn}\simeq0.5$ MeV and 
the Plank mass $M\simeq10^{19}$ GeV, we have for $\alpha\simeq1/137$
\be
m_{electron}\simeq M_{Plank}\exp(-0.375/\alpha),
\ee
involving the constant $c\simeq0.375$ which is of the same order than $2\pi/(4+\zeta)$ in eq.(\ref{mdyn}), 
and shows that this dynamical mass generation mechanism is realistic phenomenologically.\\
{\it (vi)} If one used $m_{dyn}$ for the fermion mass in the one-loop vertex calculated from the Lagrangian (\ref{bare}),
which is also regularized by $M$, the dominant contribution will be proportional to $\ln(M/m_{dyn})=c/\alpha$.
As a consequence, the limit $M\to\infty$ will be finite without redefinition of the bare coupling. The reason for this is 
that the corresponding ``one-loop'' vertex would actually consist in a partial resummation of graphs, 
since it would contain $m_{dyn}$, which purely arises from quantum fluctuations.

Finally, further studies related to the present one should aim at solving the gauge-dependence of the dynamical mass, 
in order to obtain a non-ambiguous value for the constant $c$ mentioned in {\it(iii)}, and extend this
approach to Quantum Chromodynamics.

\vspace{0.5cm}

\nin{\bf Acknowledgments} This work is partly supported by the Royal Society, UK.

\end{document}